\documentclass[draft,ef,sw]{AGUTeX}

\usepackage[dvips]{graphicx}
\usepackage{morefloats}
\usepackage[table]{xcolor}
\usepackage{multirow}

\definecolor{Red}{rgb}{1,0,0}

\setkeys{Gin}{draft=false}

\authorrunninghead{FACSK{\'O} ET. Al.}

\titlerunninghead{ONE YEAR GLOBAL MHD SIMULATION}

\authoraddr{Corresponding author: G.~Facsk{\'o}, Geodetic and Geophysical Institute,  Research Centre for Astronomy and Earth Sciences, Hungarian Academy of Sciences, H$-$9401~Sopron P.~O.~Box~5, Hungary, (facsko.gabor@csfk.mta.hu)}

\begin{document}

\title{One year in the Earth's magnetosphere: A  global MHD simulation and spacecraft measurements}

\author{G.~Facsk{\'o}\altaffilmark{1,2}, I.~Honkonen\altaffilmark{2}\thanks{Now at NASA Goddard Space Flight Center, Greenbelt, Maryland, USA}, T.~{\v{Z}}ivkovi{\'c}\altaffilmark{3,4}, L.~Palin\altaffilmark{3}, E.~Kallio\altaffilmark{5}, K.~{\AA}gren\altaffilmark{3}, H.~Opgenoorth\altaffilmark{3}, E.~I.~Tanskanen\altaffilmark{2}, S.~E.~Milan\altaffilmark{6}}

\altaffiltext{1}{Geodetic and Geophysical Institute, Research Centre for Astronomy and Earth Sciences, Hungarian Academy of Sciences, Sopron, Hungary}
\altaffiltext{2}{Finnish Meteorological Institute, P.~O.~BOX~503, FI-00101~Helsinki, Finland}
\altaffiltext{3}{Swedish Institute of Space Physics, Uppsala, Sweden}
\altaffiltext{4}{Research and Innovation, DNV GL H{\o}vik}
\altaffiltext{5}{Aalto University, School of Electrical Engineering, Espoo, Finland}
\altaffiltext{6}{Department of Physics and Astronomy, University of Leicester, Leicester, UK}

\begin{abstract}
The response of the Earth's magnetosphere to changing solar wind conditions are studied with a 3D Magnetohydrodynamic (MHD) model. One full year (155 Cluster orbits) of the Earth's magnetosphere is simulated using Grand Unified Magnetosphere Ionosphere Coupling simulation (GUMICS$-$4) magnetohydrodynamic code. Real solar wind measurements are given to the code as input to create the longest lasting global magnetohydrodynamics simulation to date. The applicability of the results of the simulation depends critically on the input parameters used in the model. Therefore, the validity and the variance of the OMNIWeb data is first investigated thoroughly using Cluster measurement close to the bow shock. The OMNIWeb and the Cluster data were found to correlate very well before the bow shock. The solar wind magnetic field and plasma parameters are not changed significantly from the $L_1$ Lagrange point to the foreshock, therefore the OMNIWeb data is appropriate input to the GUMICS$-$4. The Cluster SC3 footprints are determined by magnetic field mapping from the simulation results and the Tsyganenko (T96) model in order to compare two methods. The determined footprints are in rather good agreement with the T96. However, it was found that the footprints agree better in the northern hemisphere than the southern one during quiet conditions. If the $B_y$ is not zero, the agreement of the GUMICS$-$4 and T96 footprint is worse in longitude in the southern hemisphere. Overall, the study implies that a 3D MHD model can increase our insight of the response of the magnetosphere to solar wind conditions. 
\end{abstract}

\begin{article}

\section{Introduction}
\label{sec:intro}

Multi-spacecraft measurements provide very limited information about the near-space environment of the Earth. Satellites collect information along their orbit, in a very small region compared to the terrestrial magnetosphere with a characteristic size of several hundred thousand kilometers. Therefore a model is necessary to understand physical processes occurring in the region that we cannot reach by observations. From a mathematical perspective, global simulations of the Solar-Terrestrial interactions are described by quite a complex system of partial differential equations. Different modelling approaches exist, one of them is the full fluid or magnetohydrodynamics (MHD) description of the magnetized fluids. Various global computer simulations have been developed which use a MHD description of plasma, for example the Lyon-Fedder-Mobarry, LFM \citep{lyon04:_lyon_fedder_mobar_lfm_mhd} code, the Open Geospace General Circulation Model, OpenGGCM \citep{raeder08:_openg_simul_themis_mission}, the Block-Adaptive-Tree-Solarwind-Roe-Upwind-Scheme, BATS-R-US \citep{powell99:_solut_adapt_upwin_schem_ideal_magnet,toth12:_adapt} and the Grand Unified Magnetosphere Ionosphere Coupling simulation, GUMICS \citep{janhunen12:_gumic_mhd}, the only global Magnetosphere-Ionosphere MHD model in Europe. These four simulations are available at the Community Coordinated Modeling Center (CCMC; http://ccmc.gsfc.nasa.gov/) hosted by the NASA Goddard Space Flight Center (GSFC). These simulations has been developed by different teams and all of them have their own strengths and weaknesses. However, every simulation faces the same challenge, namely how realistic the results are. Therefore, all simulation modes have to be verified and validated by comparing the simulations to spacecraft and to ground based measurements; as well as to results obtained from other simulations. 

The Geospace Environment Modeling (GEM) ''Metric and validation'' Focus Group has been central in coordinating this activity and it has suggested performing different simulation models on selected several hour long intervals for comparison with the ground$-$based and spacecraft measurements.  The 2008-2009 GEM Metrics Challenge requested various simulation groups to submit results for four geomagnetic storm events and five different types of observations that can be modelled using the magnetosphere-ionosphere system. To compare each of the models with the observations, one hour of averaged model data was used with the Dst index, and direct comparison one minute model data with the one minute Dst index was made. Generally speaking, the empirical models provided realistic results. It has been proposed by \citep{glocer13:_crcm_bats_r_us} that MHD models of the magnetosphere could produce more realistic results if the inner magnetosphere region contained a ring current model such as the Rice Convection Model (RCM) or the Comprehensive Ring Current Model (CRCM) as they exist in SWMF, LFM and OpenGGCM. The capability of the models to reproduce observed ground magnetic field fluctuations and geomagnetically induced current (GIC) phenomenon is also an important question regarding the MHD models \citep{pulkkinen11:_geosp_envir_model_chall,pulkkinen13:_commun}, as is the validity of the models in the magnetosphere domain. The magnetic field near the geosynchronous orbit was also compared in various models. \citet{rastaetter11:_geosp_envir_model_chall} found that the empirical models perform well during weak storms, while the MHD models gave more realistic results during strong storms. If the inner magnetosphere module of the code coupled to the MHD code contained kinetic physics, the result was even closer to reality \citep{rastaetter11:_geosp_envir_model_chall}. 

Also other results obtained from MHD models have been investigated. For example, \citet{tanskanen05:_energ_augus} compared energy input and ionospheric energy dissipation from the GUMICS$-$4 simulation and data. About an order of magnitude difference was found in the energy dissipation. However, the time variation of the joule dissipation was similar in the data and in the simulation.

More recently, in \citet{honkonen13:_earth} the predictions of the BATS-R-US, the GUMICS, the LFM and the OpenGGCM were compared with the measurements of the Cluster \citep{escoubet01:_introd_clust}, the WIND \citep{acuna95:_global_geosp_scien_progr_its_inves} and the GEOTAIL \citep{nishida94:_geotail} missons; as well as the Super Dual Auroral Radar Network \citep[SuperDARN,][]{greenwald95:_darn_super} cross polar cap potential (CPCP). The most realistic simulation result near the geosynchronous orbit was found on the day side for all models outside the geosynchronous orbit. In the magnetotail, at $-130\,R_E$, simulations succeeded in reproducing well the $B_z$ component but not the $B_y$ component. The LFM magnetopause was found to be well in agreement with the empirical models. Furthermore, the BATS-R-US and the GUMICS produces a similar magnetopause but their magnetopauses were shifted in respect to the empirical models. It was also found that the OpenGGCM magnetopause varied significantly and its deviation from the empirical model was the highest \citep{honkonen13:_earth}. Overall, the magnetopause determination in MHD models was found to be a challenging task \citep{palmroth03:_storm_mhd}.

Moreover, also long duration runs have been compared with observations. \citet{guild08:_geotail_lfm} provided a two month long simulation to compare its average properties to six years of Geotail (http://www.stp.isas.jaxa.jp/geotail/) observations using the Lyon-Fedder-Mobarry global simulation model. The CPCP, the field-aligned current (FAC), downward Poynting flux and the vorticity of ionospheric convection were compared with observed statistical averages. It was shown that the LFM model produces reasonably accurate average distributions of the currents. However, the CPCP was found to be greater than the observed results. The ionospheric convention pattern was instead realistic.  Furthermore, the ionospheric field-aligned vorticity average was found to agree well with the measurements on the day side. On the other hand, the LFM model simulation used unrealistically small ionospheric conductance on the night side, and the night side vorticity was higher than observed \citep{zhang11:_lyon_fedder_mobar_mhd}.

In this paper a global MHD simulation lasting approximately one year is performed using the GUMICS-4 code with about one year of OMNIWeb data from January 29, 2002 to February 2, 2003 given as input. The structure of this paper is as follows. Section~\ref{sec:sim} presents how the year$-$long simulation was launched. Section~\ref{sec:t96} gives comparisons between the simulations and observations. Results of the comparison are discussed in Section~\ref{sec:disc}. Finally, Section~\ref{sec:concl} contains the conclusions.

\section{Simulations}
\label{sec:sim}

\subsection{GUMICS-4 model}
\label{sec:gumics4}

The Grand Unified Magnetosphere Ionosphere Coupling simulation (GUMICS, version 4) is a global simulation of the terrestrial plasma environment. The only time-dependent input parameters are the properties of the solar wind. The simulation box is $+32\,R_E$ to $-224\,R_E$ in the GSE (Geocentric Solar Ecliptic) X direction and $\pm64\,R_E$ in the Y and Z directions. Outflow conditions are applied at all boundaries of the simulation box except at the sunward wall, where the values are solar wind parameters. There are two simulation domains: the ionospheric domain at 110\,km altitude and the magnetospheric domain with an inner boundary at $3.7\,R_E$. These domains are coupled to each other, and the ionospheric potential is updated every four seconds in the simulation. The field-aligned currents (FAC) are derived from currents at the inner boundary at $3.7\,R_E$ and mapped along dipole field lines to the ionosphere. In the electrostatic ionospheric domain the Pedersen and Hall conductivities are computed from the electron precipitation and solar EUV radiation. The electrostatic potential is calculated from the conductivities and the field-aligned currents and mapped back to the magnetosphere. The electrostatic potential is mapped back along dipole field lines to the $\left(3.7\,R_E\right)$ inner boundary and applied as a convection pattern \citep[see][and references therein]{janhunen12:_gumic_mhd}. The GUMICS$-$4 grid is adaptively refined where interesting physical features occur in the si\-mu\-la\-ti\-on. The finest resolution of $1/4\,R_E$ occurs along the dayside magnetopause and near the $3.7\,R_E$ inner boundary and the coarsest resolution of $2\,R_E$ is found in the solar wind and in far down-tail regions.

Previously, a large number of synthetic simulations (the so-called GUMICS run library) were used to verify the capability of the GUMICS-4 global MHD simulation \citep{gordeev13:_verif_gumic_mhd}. These simulations, based on typical solar wind parameters, can be used as a library. These results are useful when the upstream solar wind parameters are not known. Moreover, synthetic runs are also important because they give a possibility to study the response of the magnetosphere to constant upstream parameters. It is important to note several issues considering challenges related to the year$-$long simulation. GUMICS-4 simulation cannot simulate time-dependent $B_x$ on the solar wind boundary. Hence the observed IMF $B_x$ cannot be used in the simulation, otherwise the simulation produces magnetic divergence and the solution becomes non-physical. There are two general ways to avoid introducing divergence of magnetic field into the simulation at the solar wind boundary. One is to set $B_x$ from the input to zero and to add a constant background value to the magnetosphere and the dipole field. When a discontinuity is present it is possible to determine the appropriate reference frame of the discontinuity, when the magnetic field divergence is zero across the boundary layer \citep{raeder03:_global_magnet_tutor}. However, this minimum variance method is applicable only for short intervals, thus, only some other method could be applied for a massive number of simulations. 

\subsection{Inputs to the model: OMNI data}
\label{sec:inputs}

\begin{figure}
\centering
\noindent\includegraphics[width=0.9\textwidth]{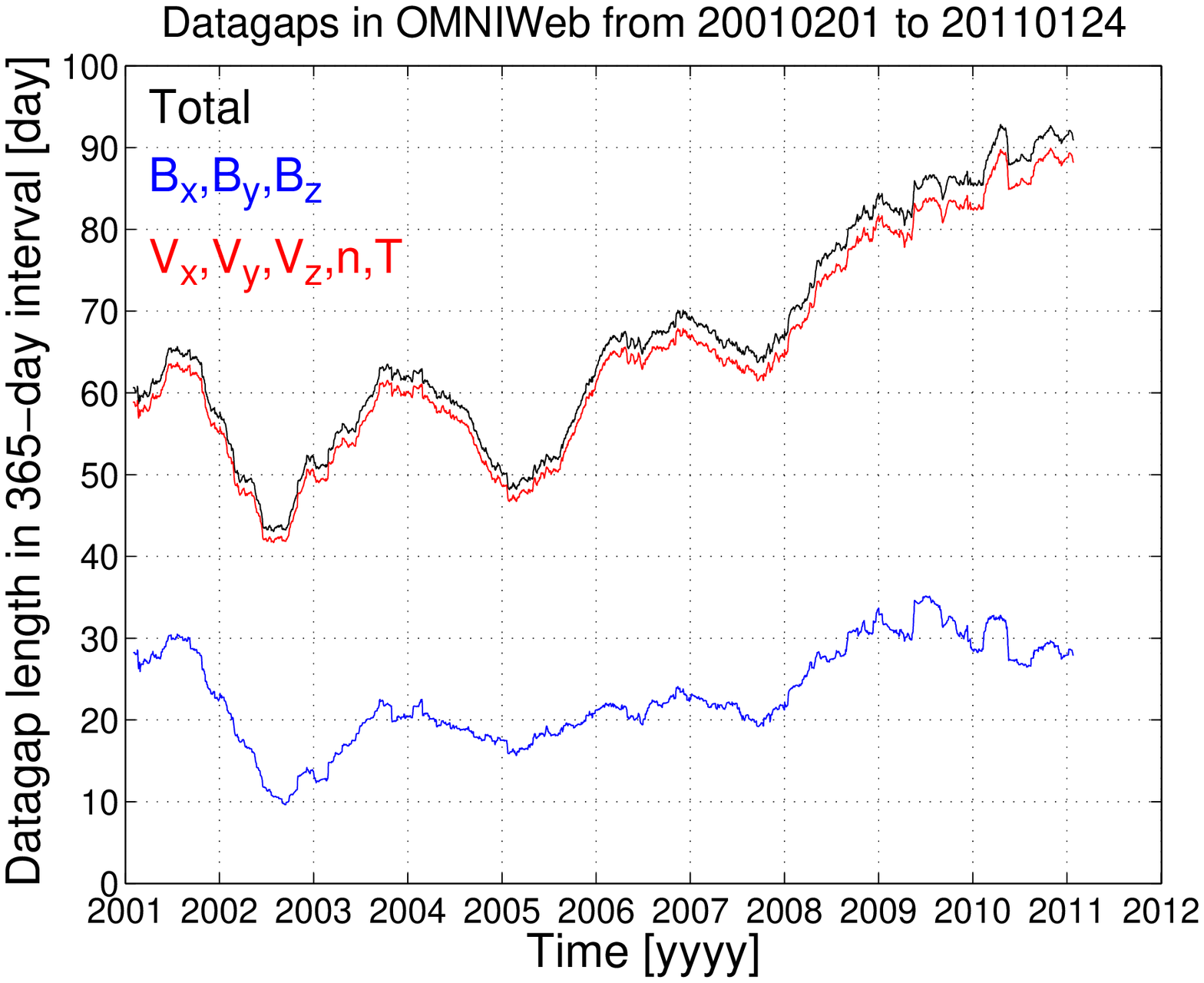}
\caption{The length of total data gap time in 365 days long sliding window using the OMNI one minute averaged solar wind magnetic field and ion plasma data from February 1, 2001 to January 24, 2011. Red: solar wind plasma measurements ($V_y, V_y, V_z$ solar wind velocity, n: solar wind density, T: solar wind temperature). Blue: interplanetary magnetic field ($B_x, B_y, B_z$). Black: the length of data gap in plasma and/or field measurements the total length of data gap in all datasets. The studied interval is the Cluster mission operation time.}
\label{fig:gaps}
\end{figure}

The one$-$year simulation was launched using OMNI (http://omniweb.gsfc.nasa.gov/) solar wind data as input to the GUMICS simulation, whose assimilation was a part of the European Cluster Assimilation Techniques (ECLAT) project. To achieve the maximum amount of dynamic simulation, the maximum amount of input data is necessary. The OMNI data contains data gaps, therefore the minimum total data gap for one year shows the optimal interval of input files for simulations. In Figure~\ref{fig:gaps} the total length of data gaps is plotted using OMNI data from the start of the Cluster mission to 365 days before the mission ended. The lengths of the data gaps are determined in each 365$-$day intervals starting from a given day. This calculation is made for plasma data (density, temperature and solar wind velocity; see Figure~\ref{fig:gaps}, red curve) and magnetometer measurements (Figure~\ref{fig:gaps}, blue curve). The length of data gap in either instrument is plotted in black. When the plasma data are missing, the magnetic field data could be useful, but when the magnetic field measurements are disturbed, the plasma data is almost always also corrupted (or the data gaps in the B and the particle data often coincide briefly). After a year long calibration period, the total length of data gaps slowly increases. The data quality of the plasma instrument decreases faster than the magnetic field data. Indeed, the total length of data gaps in the magnetic field measurements is almost constant between 2003 and 2008. The total data gap has three local minimums following the three visible minimums of plasma data gaps, but the OMNI data has the shortest data gap length in 2002 and 2003. It should be noted that both the ion plasma and magnetic field data are necessary for the simulations as input parameters. Based on the analysis above, the interval from February 1, 2002 to January 31, 2003 is selected to be simulated. 

As mentioned before, the GUMICS-4 is the only 3D MHD model in Europe which contains magnetosphere-ionosphere coupling. In this study the GUMICS-4 results are compared and validated with the Cluster mission.  The Cluster-II mission was launched in July and August of 2000 and it consists of four similar spacecraft, equipped with eleven instruments aboard \citep{escoubet01:_introd_clust}. The four probe forms a tetrahedron and their orbit is an almost polar 57\,h long elliptical orbit with 19000\,km perigee and 119000\,km apogee. The orbit crosses the magnetosphere, the magnetosheath, the foreshock, the bow shock and the magnetotail. The special formation of the four spacecraft allows the study of these plasma regions using multi-spacecraft methods. In this study we use FluxGate Magnetometer (FGM) magnetic field data \citep{balogh01:_clust_magnet_field_inves} and Cluster ion spectrometry (CIS) ion plasma data \citep{reme01:_first_earth_clust_cis} for comparison.

It is worth noting the following challenges associated with the computational perforce of the used simulation model. The GUMICS$-$4 model has not yet been parallelized and runs much slower than real time at the resolution required for this study. Hence a 1$-$year simulation would take decades to complete. Therefore, the 1$-$year time interval was broken up into 57\,hour intervals to coincide with full Cluster orbits. This method enabled easy comparison between the simulated results and the observations made by Cluster. In practice, the simulated time period had to be longer that 57\,hours because the GUMICS-4 needs at least one hour input data as initialization. 

It is also important to note the following practical issue concerning the simulations presented in this study. Approximately one year (368 days, 155 Cluster orbits) was selected and given as input to the GUMICS-4. The simulation of 57\,hours$-$long orbits would have completed within a half year, because the GUMICS is 72 times slower than real time on the Cray supercomputer of the Finnish Meteorological Institute (FMI). Subdividing each orbit into 12\,segments allowed us to complete each orbit in as little as 18\,days and to start implementing post-processing and analysis procedures. Up to 480\,segments (out of the total of 1860) were run in parallel, allowing us to complete all calculations in less than 5\,months (including computer down time and other operational delays). Each Cluster orbit was divided into twelve 4.75\,hours long slices with one hour initialization period to parallelize the simulation. The initialization was done using one hour constants of solar wind input values. One minute resolution OMNI data was used as solar wind input. As it could have been seen on Figure~\ref{fig:gaps}, the OMNI contained significant amount of data gaps. The data gaps were filled using linear interpolation between the last valid data before the data gap and the first valid data after the data gap. The magnetic field is treated in the following way: the $B_x$ component of the OMNI magnetic field was not used and replaced with its average added to the background magnetic field during each interval. Moreover, the average dipole tilt angle was used for each slice. The magnetospheric and ionospheric results were always saved once in every five minutes. 

\subsection{Timeshift}
\label{sec:timeshift}

As already mentioned, the solar wind input files of the GUMICS$-$4 simulation were one minute of resolution OMNI solar wind data. The OMNI shifted its data to the subsolar point of the terrestrial bow shock (http://omniweb.gsfc.nasa.gov/html/HROdocum.html\#3). However the inbound wall of the GUMICS$-$4 simulation was at +32\,$R_E$ in the GSE X direction. Thus, the OMNI input files should have been shifted to the +32\,$R_E$ boundary. Applying the reverse delay from the bow shock to the +32\,$R_E$ GSE X, the time shift should have been done on a case by case basis using the method we describe below.

\begin{figure}
\centering
\noindent\includegraphics[width=0.95\textwidth]{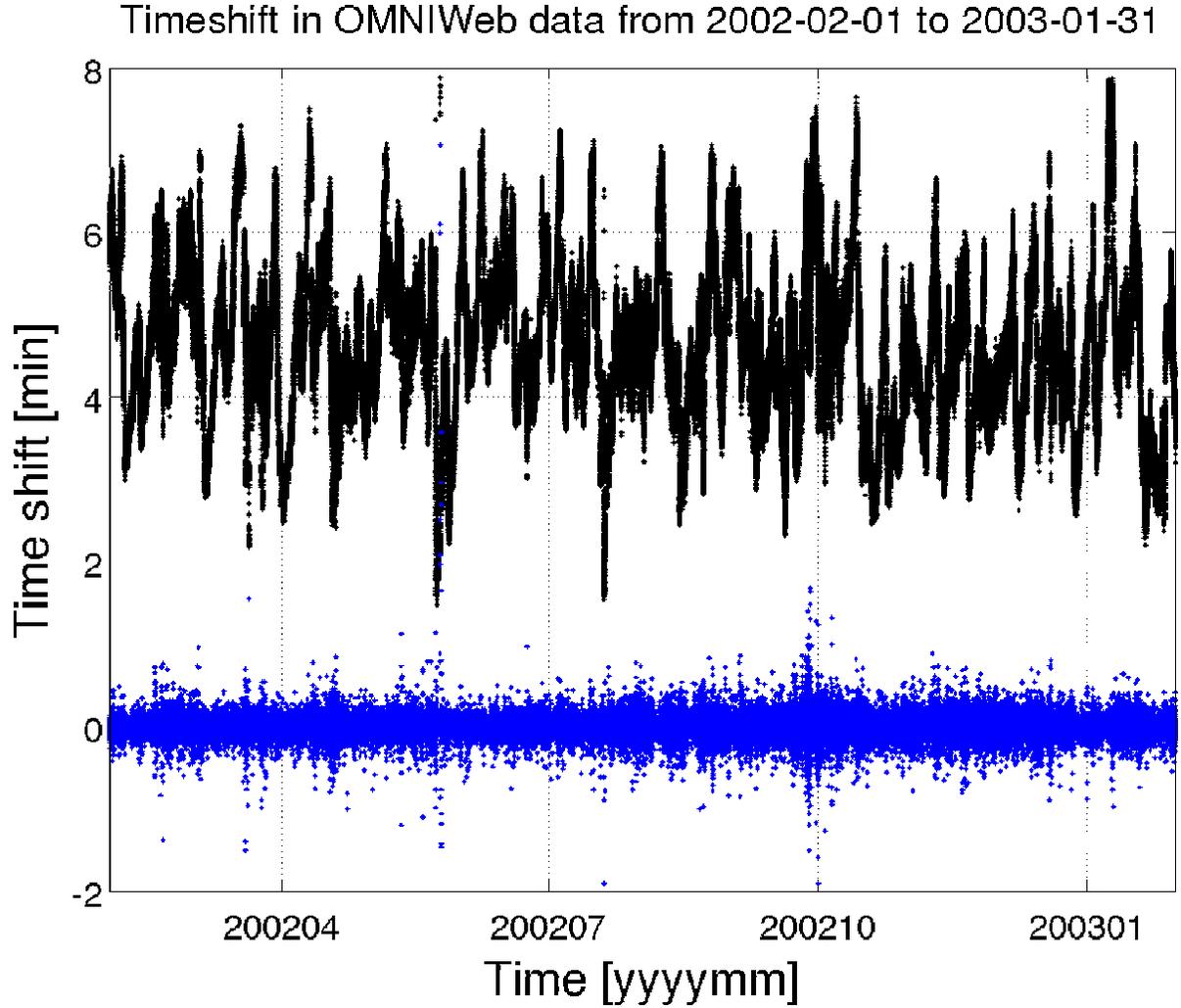}
\caption{Black: The timeshift of the solar wind from the subsolar point of the terrestrial bow shock and the +32\,$R_E$ (the ingoing wall of the GUMICS$-$4 simulation box) calculated using the OMNI one minute averaged solar wind magnetic field and ion plasma data from February 1, 2001 to January 31, 2003. Blue: the difference of the timeshift was computed every 5th minutes.}
\label{fig:shift}
\end{figure}

The OMNI calculated the magnetopause position using the \citet{shue97} model. The bow shock position was calculated using the \citet{farris94:_deter} bow shock model based on the above described magnetopause model. The GUMICS$-$4 inbound wall was always at +32\,$R_E$, thus using the solar wind speed GSE X component, the time shift relative to the subsolar point of the bow shock could have always been easily calculated (Figure~\ref{fig:shift}, black dots). On the plotted time series it was visible that the time shift was roughly between +2 and +8\,minutes and the average was around +4--5\,minutes. The dynamic simulation results were saved at every five minutes, thus the timeshift of the simulation parameters was only one input file value point or less. Note that Figure~\ref{fig:shift} showed one minute of resolution values because it was derived from one minute resolution OMNI data. The difference of the timeshift of the saved data (every 5th timeshifts) was mostly zero minutes (Figure~\ref{fig:shift}, blue dots). 

\subsection{Quality of the solar wind inputs}
\label{sec:input}

\begin{figure}
\centering
\noindent\includegraphics[width=0.9\textwidth]{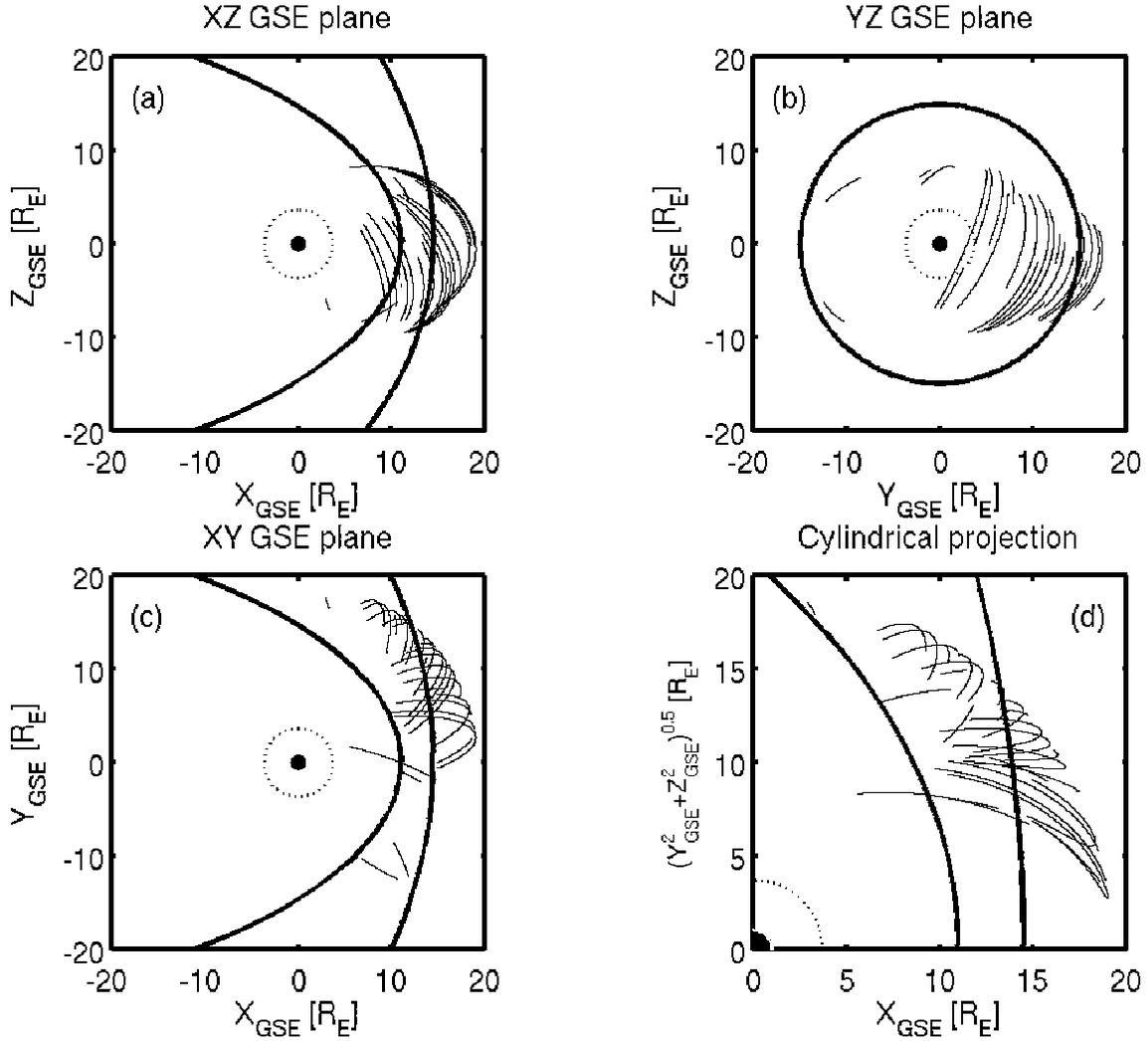}
\caption{The Cluster reference spacecraft orbits plotted in the intervals which are listed in Table~\ref{tab:omniclcorr} in the GSE system. Average bow-shock and magnetopause positions are drawn on all plots \citep[][respectively]{peredo95:_three_alfven_mach,tsyganenko95:_model_earth}. The black dots at $3.7\,R_E$ show the boundary of the GUMICS$-$4 inner magnetospheric domain. The black circle in the origo of all plots shows the size of the Earth. The four panels show the same orbits presented in (a) the XY-plane, (b) the YZ-plane, (c) the XY-plane and (d) a cylindrical projection.}
\label{fig:orbits}
\end{figure}

\begin{figure}
\centering
\noindent\includegraphics[width=0.85\textwidth]{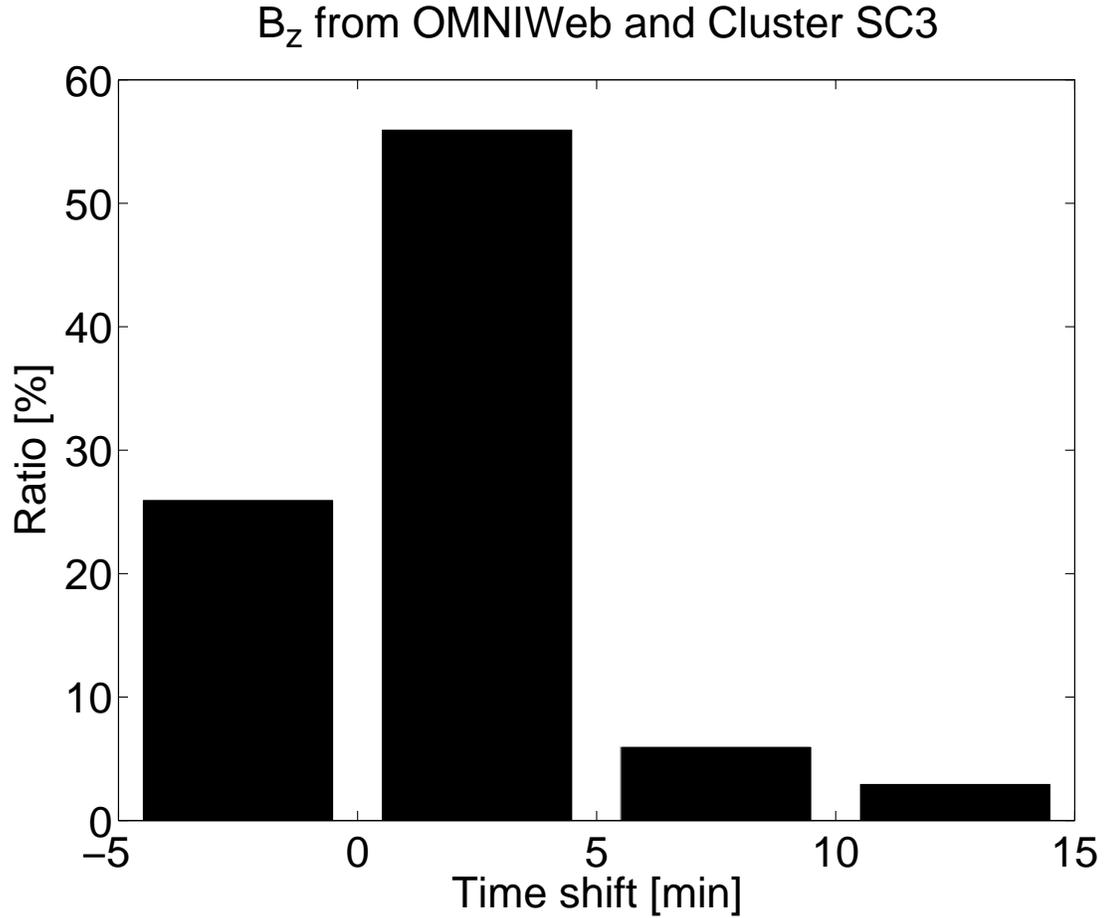}
\caption{The histogram of the calculated timeshift using cross correlation of the one minute resolution OMNI and one minute averaged Cluster SC3 magnetic field $B_z$ component data. The distributions of timeshift in minutes from Table~\ref{tab:omniclcorr}. Each column gives the relative ratio of the number of the timeshift between the indicated lower and higher values of the bar.}
\label{fig:corrhist}
\end{figure}

The quality of the simulation result depends on the quality of the input solar wind values and, therefore, it is important to note the following issues about the adopted inputs. As mentioned before, the OMNI data is created from various spacecraft measurements: ACE \citep{chiu98:_ace_spacec}, WIND and IMP 8 (http://omniweb.gsfc.nasa.gov/html/omni\_min\_data.html\#1a). The solar wind parameters are shifted to the subsolar point of the terrestrial bow shock. There are at least two uncertainties to this method: the position of the subsolar point and the quality of the data created. To test the quality of the input data we selected time intervals of several hours durations in the magnetic field measurements of the Cluster reference spacecraft (SC3) when the SC3 were situated in the solar wind (Table~\ref{tab:omniclcorr}, Figure~\ref{fig:orbits}). The selection of the solar wind intervals was made manually. The bow shock crossings were visible as a large jump of the magnetic field magnitude from high ($\sim$25\,nT) to lower value ($\sim$5\,nT) and the solar wind speed increased from $\sim$100-200\,km/s to $\sim$400-800\,km/s. At the same time the density of the plasma also decreased. The same intervals are also selected in the OMNI magnetic field data. One minute averaged data is created from spin resolution Cluster SC3 magnetic field $B_z$ component measurements. Data gaps are filled by linear interpolation. Cross correlations with and without time shifts are calculated between $B_z$ observation data and model results. Time shifts that maximize correlations are listed in Table~\ref{tab:omniclcorr}. The correlation is good between the different time series. The coefficient values are greater than 0.8, thus the shape of the curves are quite alike. Note that 80\,\% of the timeshifts are less than five minutes (Figure~\ref{fig:corrhist}) and 2/3 of them are less than 2 minutes (not shown). For a comparison, the solar wind moves typically during that time only $\sim1\,R_E$, which is not a significant distance in a global scale. Some of these large timeshifts could be explained by very disturbed magnetic field (Cluster was at the quasi-parallel foreshock). Note that large timeshifts are related to the long data gap. 

\subsection{Continuity of simulation results}
\label{sec:jumps}

\begin{figure}
\centering
\noindent\includegraphics[width=0.7\textwidth]{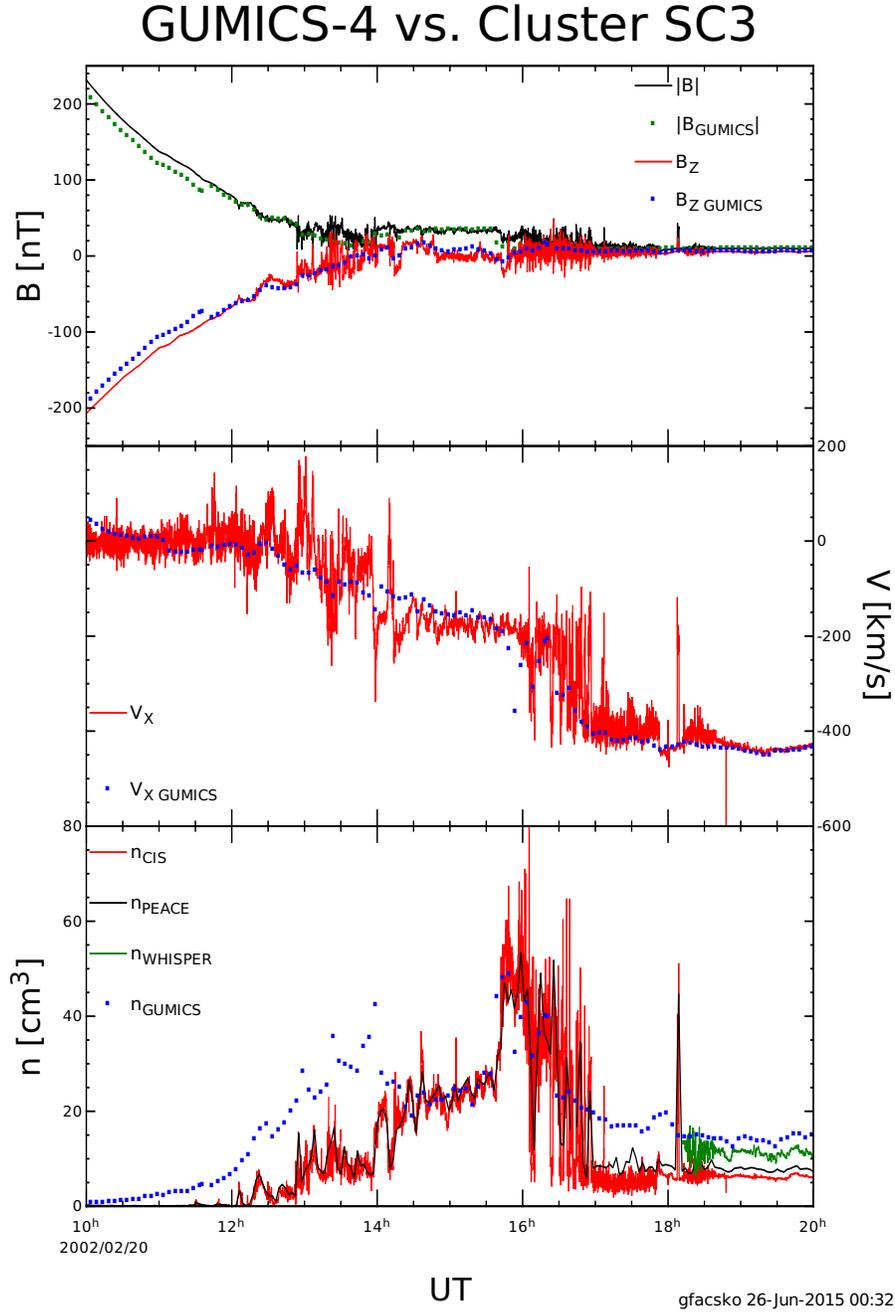}
\caption{Comparison of Cluster SC3 measurements and GUMICS$-$4 simulation results along the Cluster reference/SC3 orbit in the simulation space from 10:00 to 20:00 (UT) on February 20, 2002. (Top) The magnitude and GSE Z component of the magnetic field. (Middle) The GSE X component of the solar wind velocity (Bottom) the ion and electron densities. In the panels the simulated values are shows by dots and the measured values by solid lines. See text for details.}
\label{fig:comp}
\end{figure}

The 368 days are simulated in 1860 slices or subintervals. This approach could be considered as one large simulation if the jumps of the parameters at the boundary of the slices were not significant compared to the fluctuations of the same parameters inside the slices. Figure~\ref{fig:comp}, the magnetic field magnitude and the $B_z$ component are plotted on the top in the GSE reference frame (this coordinate system is also used below). The simulation results are represented by dots and measurements by solid lines. The temporal resolution of the simulations is five minutes while in Figure~\ref{fig:comp} the original Cluster data has 4\,s resolution. Both the values and the shape of the curves correspond well for all magnetic field components. As can be seen in the middle of Figure~\ref{fig:comp}, the simulated solar wind velocity X component is in good agreement with observations, as well as the velocity Y component. On Figure~\ref{fig:comp}, bottom panel, the simulated ion density is plotted together with the observed CIS HIA ion density \citep{reme01:_first_earth_clust_cis} and the WHISPER \citep{decreau01:_early_whisp_clust,trotignon10:_whisp_relax_sound_clust_activ_archiv} and PEACE electron densities \citep{johnstone97:_peace,fazakerley10:_peace_data_clust_activ_archiv,fazakerley10:_clust_peace_in_calib_status}. Note that the simulated and observed plasma densities behave similarly although detailed values differ. Note also that the MHD simulation results are closer to the WHISPER electron density. The change of the orbit number, and sometimes the border of the simulated slices, makes non-physical jumps in the parameters, because of the different tilt angles and $B_x$ average given to the run. It should be noted that the magnetic field components correspond very well in values and shape, including the $B_x$ component, that is changed to an average value for each slide. The plotted interval in Figure~\ref{fig:comp} contains two borders of slices at 11:38 and 16:23 on February 20, 2002. The border at 11:38 is visible in the magnetic field magnitude at the top of Figure~\ref{fig:comp}; the other cannot be seen because of the short data gap at the boundary. The jump in the plasma density is usually smaller than the variance of the slices before and after the boundary (not shown here).

\begin{figure}
\centering
\noindent\includegraphics[width=0.8\textwidth]{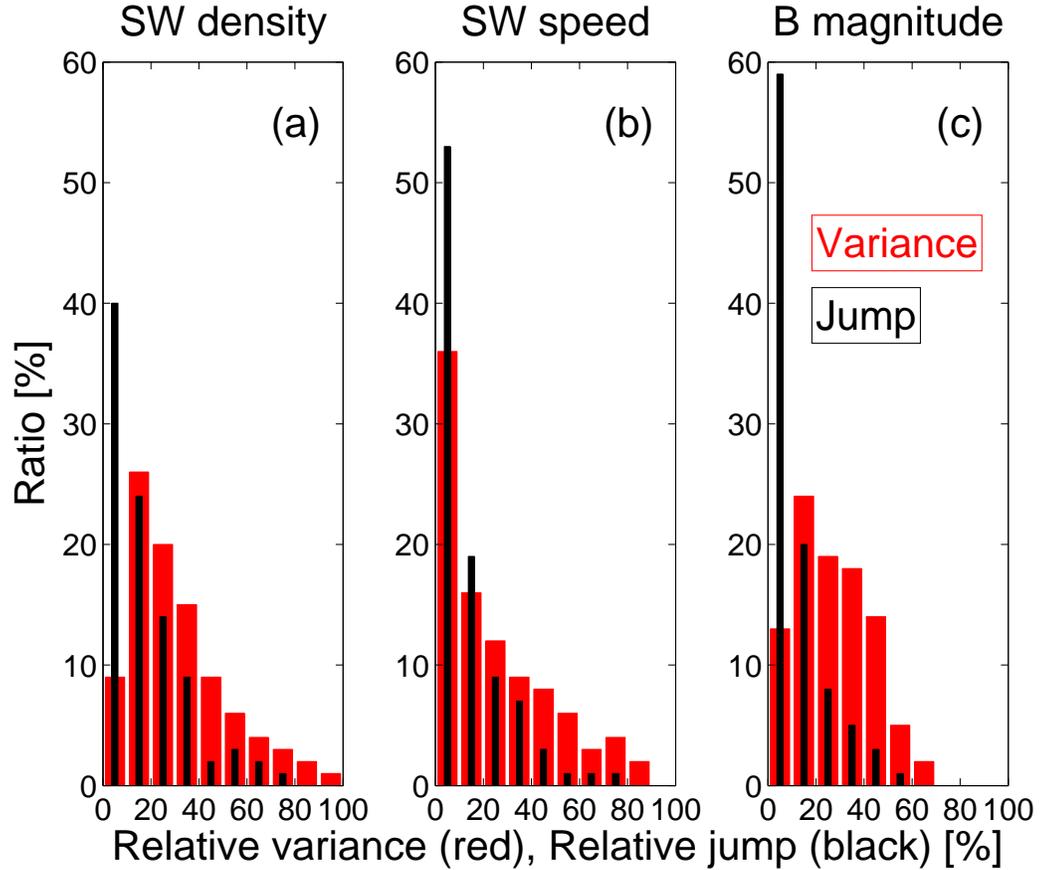}
\caption{Comparison of changes of plasma parameters between the simulation sub-intervals (black bars) and the variance (red) of solar wind density (a), velocity (b) and magnetic field magnitude (c). Both the jump and the variance are relative, the quantities were divided by the mean value of the previous slice. The numbers of the distributions were normalized by the sum of the amounts. All quantities are unitless, given in percent.}
\label{fig:compstat}
\end{figure}

In Figure~\ref{fig:compstat} the distribution of the jumps between slices along the Cluster SC3 orbit, the last value of the previous and the first value of the following slice, is drawn. The status of the simulations is saved every five minutes, therefore, it is not useful to compare the mean of a short interval before and after the boundary and it does not provide very different results (not shown). The deviation (or difference) of the solar wind density, velocity and magnetic field magnitude is divided with the mean value of the previous slice (Figure~\ref{fig:compstat}, black bars). This is what defines a jump in the following explanation. The variance of each quantity is normalized in respect to the mean density, velocity and magnetic field, respectively (Figure~\ref{fig:compstat}, red). The distributions of the relative variance and the relative jump were normalized by the sum of the distribution. As can be seen in Figure~\ref{fig:compstat}, the relative jump distribution of all quantities tends towards the smaller values. 65-75-80\,\% of the density, velocity and magnetic field relative jump is less than 20\,\%. Moreover, the relative variance has less steep distribution on all plots, these values are higher than the relative jump on all plots - except the interval of the lesser values. Note also that the maxima of the density and the magnetic field relative variance distributions are at higher values than the relative jump. The relative jump is also usually smaller than the relative natural fluctuation of the previous sub-interval or slice. Therefore, the slightly different values of the last and the first points cannot be considered as a serious break between the simulated intervals because these relative jumps are comparable to the normal variation of the slices.

\section{Comparison of Cluster C3 footprints from the T96 and GUMICS models}
\label{sec:t96}

In this section basic features of the output data to GUMICS and its results are presented. The simulated data made it possible to make numerous type of analysis and only part of them are presented in this paper. More detailed analysis of various and different aspects of the results has already been published for instance in \citet{juusola14:_statis_gumic_mhd} and \citet{kallio15:_proper}. Here we provide a different analysis and its results.

The footprint determination is given in Section~\ref{sec:t96map}. A comprehensive investigation of the GUMICS$-$4 magnetic field mapping capability is given in Section~\ref{sec:t96results}. This gives a possibility to validate GUMICS$-$4 using empirical formulas as has been done previously in \citet{gordeev13:_verif_gumic_mhd}. Moreover this model-model comparison is also a check of the magnetic field mapping based on T96, because the accuracy of the Tsyganenko method has never been studied previously.

\subsection{Magnetic field mapping in GUMICS-4 global MHD code}
\label{sec:t96map}

\begin{figure}[htb]
\centering
\includegraphics[width=0.7\textwidth]{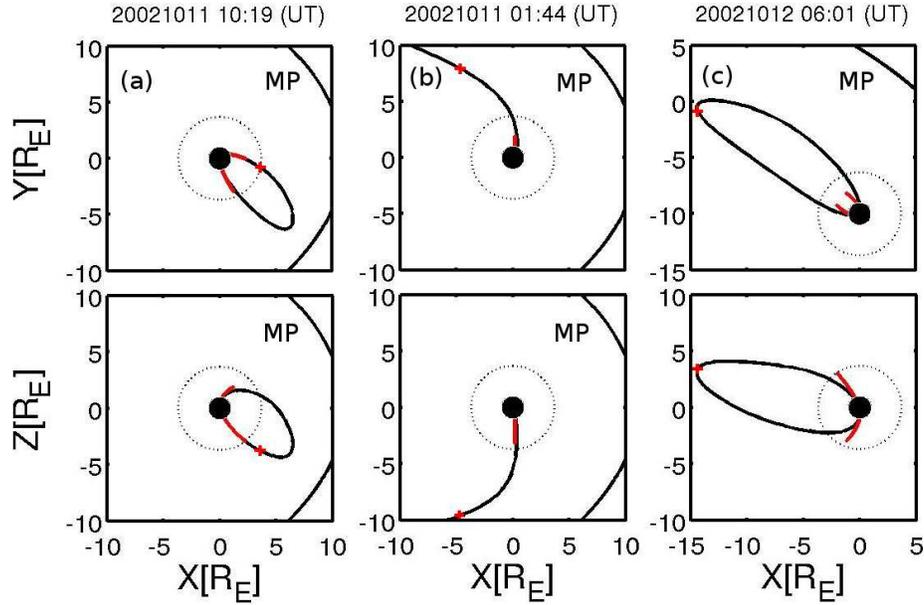}
\caption{Comparison of Cluster SC3 (+) magnetic footprints determined from the GUMICS$-$4 simulations (red dots) and the T96 model (black line). The position of the magnetopause is based on the \citet{tsyganenko95:_model_earth} model. (Only the first and the last positions of the GUMICS$-$4 magnetic field mapping are shown in the magnetospheric domain.) The red crosses mark the Cluster SC3 location. The magnetic field mapping method starts at the Cluster SC3 position in the magnetosphere domain of the simulation. The black dots at $3.7\,R_E$ show the boundary of the GUMICS$-$4 ionospheric domain. The reference frame is GSE in all figures. The black circle in the origo of all plots depicts the Earth. (a) Example of a closed field line case when the virtual Cluster SC3 is in the terrestrial magnetosphere simulated by the GUMICS$-$4. (b) Example of an open-closed field line, when Cluster is magnetically connected to the magnetosphere. (c) Example of closed field lines, when the Cluster reference spacecraft is located in the nightside magnetopshere. Note that the difference between the GUMICS-4 and T96 is higher in case (c) than in cases (a) and (b).} 
\label{fig:mapping}
\end{figure}

In the magnetic field mapping the spacecraft location is projected from the magnetosphere to the ionosphere along the magnetic field lines. Therefore, the footprint of the spacecraft is the magnetic conjugate to the spacecraft location through the same field line. Depending on whether the spacecraft is outside or inside the magnetosphere on an open (lobe field line) or a closed field line, there will be 0, 1 or 2 (one per hemisphere) footprints found, respectively. In Figure~\ref{fig:mapping}, black solid lines lead from the spacecraft location to the top of the ionosphere.

Based on the geopack documentation the T96 model uses its own empirical magnetic field in the magnetosphere until a certain specific distance below which the International Geomagnetic Reference Field (IGRF) is used. The IGRF is the empirical representation of the Earth's magnetic field recommended for scientific use by the Working Group of the International Association of Geomagnetism and Aeronomy (IAGA). The IGRF model represents the main (core) field without external sources based on all the available data sources including geomagnetic measurements from observatories, ships, aircrafts and satellites \citep[][and references therein]{tsyganenko95:_model_earth,tsyganenko96:_model_birkel}.  The IGRF model is applied below $3.7\,R_E$, which is the GUMICS$-$4 domain boundary distance for comparison. Tsyganenko's geopack is used here for visualization and an approximate validation of the GUMICS$-$4 footprint determination results. The comparison is from the footprint data based on the T96 model created by the contributors of the St.~Petersburg State University [private communication]. 

The GUMICS$-$4 uses the GSE reference frame. Its own magnetic field line trace tool \citep[see][]{janhunen12:_gumic_mhd} determines the coordinate in GSE where the magnetic field line starts from where the spacecraft location crosses the boundary of the magnetospheric and ionospheric domains at $3.7\,R_E$ (Figure~\ref{fig:mapping}). The red dots in Figure~\ref{fig:mapping} show the spacecraft locations and, on the domain boundary are the start and the end of the field line tracing. The tool does not share the steps of the magnetic field mapping. The magnetic field from the inner magnetosphere boundary to the ionosphere is mapped along a dipole field. First all locations are transformed to the Solar Magnetic (SM) system because the SM system is the reference frame of the tilted terrestrial magnetic field, thus it is easier to continue the field line mapping in the SM system. The magnetic field was assumed to be dipolar:
\begin{eqnarray}
\label{eq:dipole}
B_x &=& \frac{-3 k_0 x z}{(x^2 + y^2 + z^2)^{\frac{5}{2}}} \\
B_y &=& \frac{-3 k_0 y z}{(x^2 + y^2 + z^2)^{\frac{5}{2}}} \\
B_z &=& \frac{-3 k_0 (z^2-\frac{(x^2+y^2+z^2)}{3})}{(x^2+y^2+z^2)^{\frac{5}{2}}}\\
l_s &=& \frac{step\_size}{(B_x^2 + B_y^2 + B_z^2)^{\mathbf{\frac{1}{2}}}},
\end{eqnarray}
where $B_x$, $B_y$, $B_z$ are the component of the dipole field and $k_0 = 8\times10^{15}\,T\,m^3$. The $step\_size$ parameter is an initial selected spatial distance based on stability and accuracy considerations. In this study the step size was 100\,km. On a closed field line, the field line tracing algorithm follows the field line in the direction of both hemisphere using the $l_s$ step. The algorithm stops at 100km altitude in the Earth ionosphere (see Figure~\ref{fig:mapping}, red dots in the ionosphere domain). On an open field line, the field line tracing is stopped at the boundary of the magnetospheric simulation box (see Section~\ref{sec:sim}). The results are converted to the GSE system and saved. These coordinates are compared to the T96 footprint coordinates in Section~\ref{sec:t96results}. 

\subsection{Results}
\label{sec:t96results}

\begin{figure}[htb]
\centering
\includegraphics[width=0.9\textwidth]{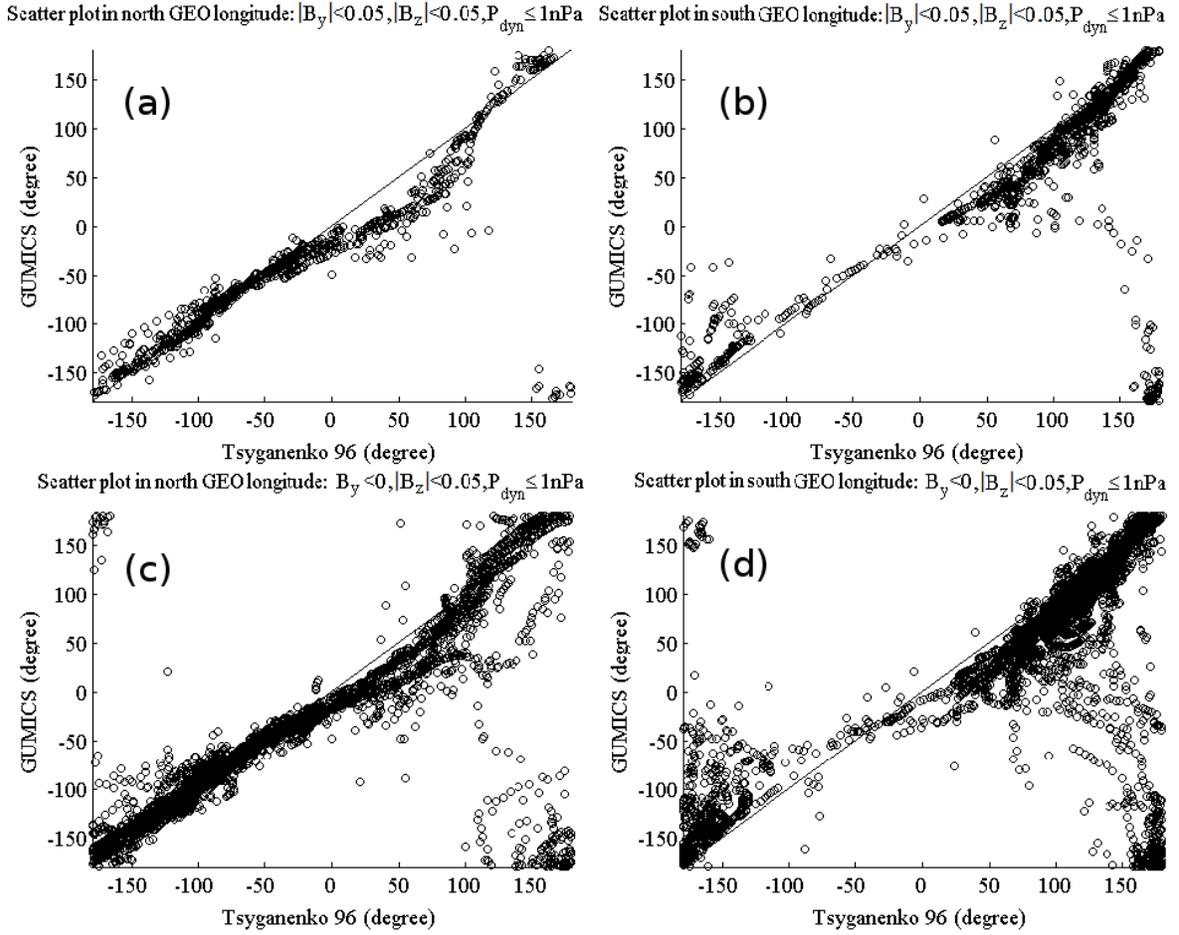}
\caption{Scatter plot of the position of the magnetic footprint in longitude for quiet conditions: (a) northern hemisphere, (b) southern hemisphere. Scatter plot of the position of the magnetic footprints in longitude when $B_{y}<0$: (c) northern hemisphere, (d) southern hemisphere.}
\label{fig:scatter}
\end{figure}

The correlation between the GUMICS$-$4 and T96 models has been investigated at different IMF magnetic field and solar wind dynamic pressure conditions. Investigations are made for each combinations of $B_{\mathrm{y}} < 0$\,nT, $B_{\mathrm{y}} > 0$\,nT and $\left|B_{\mathrm{y}}\right| < 0.05$\,nT, $B_{\mathrm{z}} < 0$\,nT, $B_{\mathrm{z}} > 0$\,nT and $\left|B_{\mathrm{z}}\right| < 0.05$\,nT, $P_{\mathrm{dyn}} < 1$\,nPa and $P_{\mathrm{dyn}} > 1$\,nPa. Figure~\ref{fig:scatter}a,\,b shows GUMICS$-$4 versus T96 footprints in geographical coordinates for the northern (panel a) and southern (panel b) hemispheres during quiet conditions, when $\left|B_{y}\right| < 0.05$\,nT and $\left|B_{z}\right| < 0.05$\,nT and when $P_{\mathrm{dyn}} < 1$\,nPa. In the northern hemisphere below $0^o$ longitude, the models are in good agreement. In the region between $0^o$ and $100^o$ both hemispheres display a deviation between the models, although the southern hemisphere footprints show a slightly better agreement between the models. In the cases when $B_{\mathrm{y}} < 0$\,nT, the correlation between the models becomes worse particularly for the southern hemisphere, c.f. Figure~\ref{fig:scatter}c,\,d. The same results are obtained for $B_y > 0$\,nT (here not shown).

The correlation in latitude does not seem to depend on magnetospheric conditions (not shown here). The footprints in the southern hemisphere, however, show less correlation. One possible explanation to this could be that GUMICS$-$4 assumes a simple dipole magnetic field for the inner magnetosphere within 3.7 \,$R_{\mathrm{E}}$ while T96 uses a more realistic intrinsic magnetic field model. For this reason, we  mostly focus on the northern hemisphere on the plots.

Furthermore, we have compared GUMICS$-$4 and T96 footprints for two principal periods: February and March, and July and August, 2002. In February and March, the perigee of the Cluster SC3 spacecraft occurred in the inner magnetosphere on the night side, whereas the apogee did not reach the solar wind. Apart from intersections with the magnetic cusp, magnetospheric conditions were relatively quiet along the Cluster orbit, as Cluster spacecraft mainly was located in the inner magnetosphere. On the contrary, in July and August, the apogee occurred in the magnetotail, giving better opportunities for the Cluster spacecraft to be exposed to substorm-related magnetospheric disturbances.

\begin{figure}
\begin{center}
\includegraphics[width=0.65\textheight,angle=0]{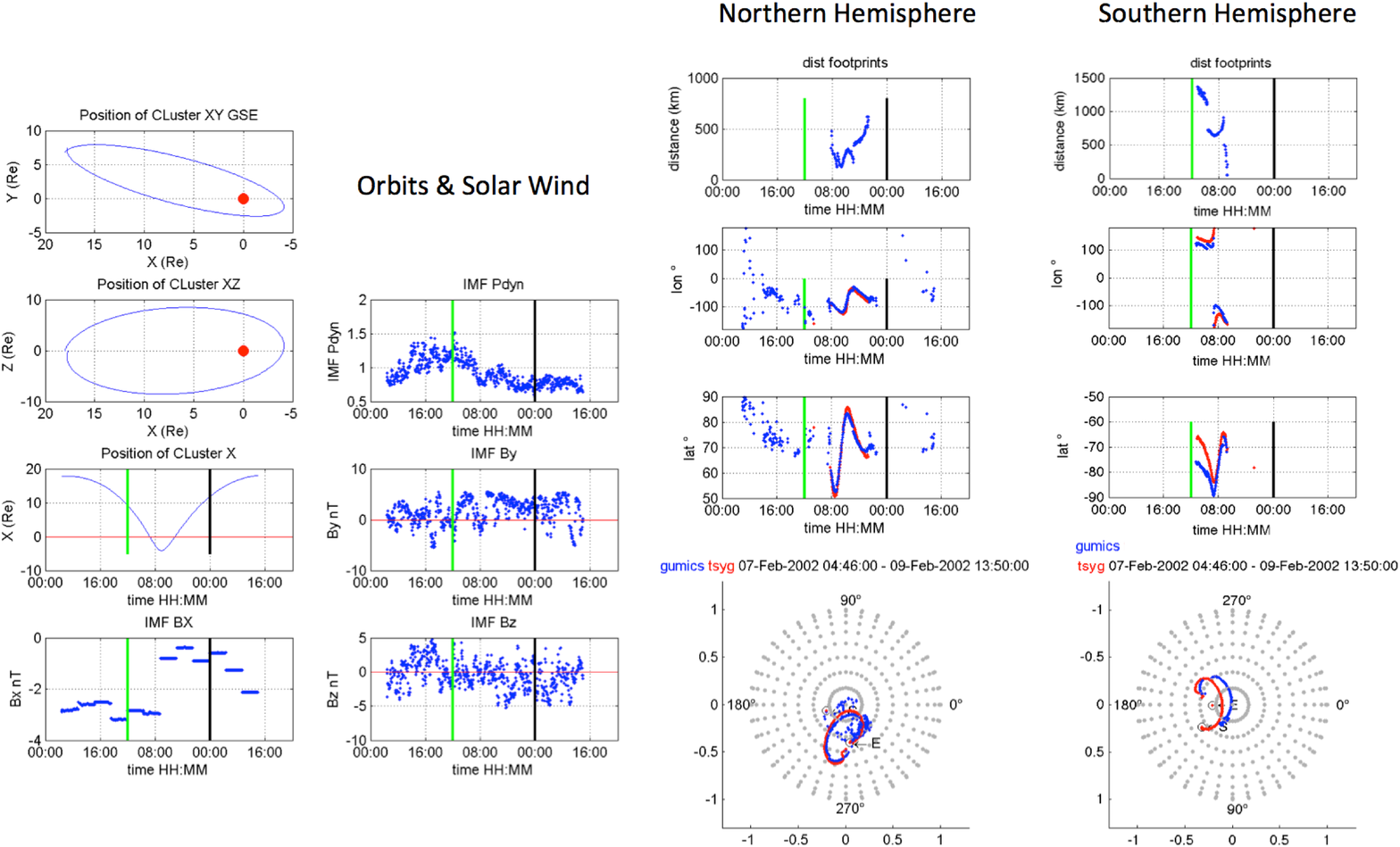}
\caption{Example for magnetic footprints analysis on February 7\textsuperscript{th}, 2002, northern and southern hemispheres. \textit{1\textsuperscript{st} column 1\textsuperscript{st} row:} the Cluster SC3 orbit in the XY GSE plane. \textit{1\textsuperscript{st} column 2\textsuperscript{nd} row:} the Cluster SC3 orbit in the XZ GSE plane. \textit{1\textsuperscript{st} column 3\textsuperscript{rd} row:} the Cluster SC3 X position. \textit{1\textsuperscript{st} column 4\textsuperscript{th} row:} the $B_x$ magnetic field GSE X component. \textit{2\textsuperscript{nd} column 2\textsuperscript{nd} row:} the solar wind dynamic pressure. \textit{2\textsuperscript{nd} column 3\textsuperscript{rd} row:} the Interplanetary Magnetic Field (IMF) GSE Y component. \textit{2\textsuperscript{nd} column 4\textsuperscript{th} row:} the IMF GSE Z component. \textit{3\textsuperscript{rd} column (northern hemisphere) 1\textsuperscript{st} row:} the distance of footprints determined from the GUMICS$-$4 simulations and T96 model. \textit{3\textsuperscript{rd} column 2\textsuperscript{nd} row:} the longitude of the T96 (red) and GUMICS$-$4 (blue) footprints in the SM system. \textit{3\textsuperscript{rd} column} 3\textsuperscript{rd} row: the latitude of the T96 (red) and GUMICS$-$4 (blue) footprints in the SM system. \textit{3\textsuperscript{rd} column 4\textsuperscript{th}-5\textsuperscript{th} row:} the T96 (red) and GUMICS$-$4 (blue) footprints in SM coordinates. \textit{4\textsuperscript{rd} column (southern hemisphere) 1\textsuperscript{st} row:} the distance of footprints determined from the GUMICS$-$4 simulations and T96 model. \textit{4\textsuperscript{rd} column 2\textsuperscript{nd} row:} the longitude of the T96 (red) and GUMICS$-$4 (blue) footprints in the SM system. \textit{4\textsuperscript{rd} column 3\textsuperscript{rd} row:} the latitude of the T96 (red) and GUMICS$-$4 (blue) footprints in the SM system. \textit{4\textsuperscript{rd} column 4\textsuperscript{th}-5\textsuperscript{th} row:} the T96 (red) and GUMICS$-$4 (blue) footprints in SM coordinates.}
\label{fig:feb7}
\end{center}
\end{figure}

\begin{figure}
\begin{center}
\includegraphics[width=0.75\textheight,angle=0]{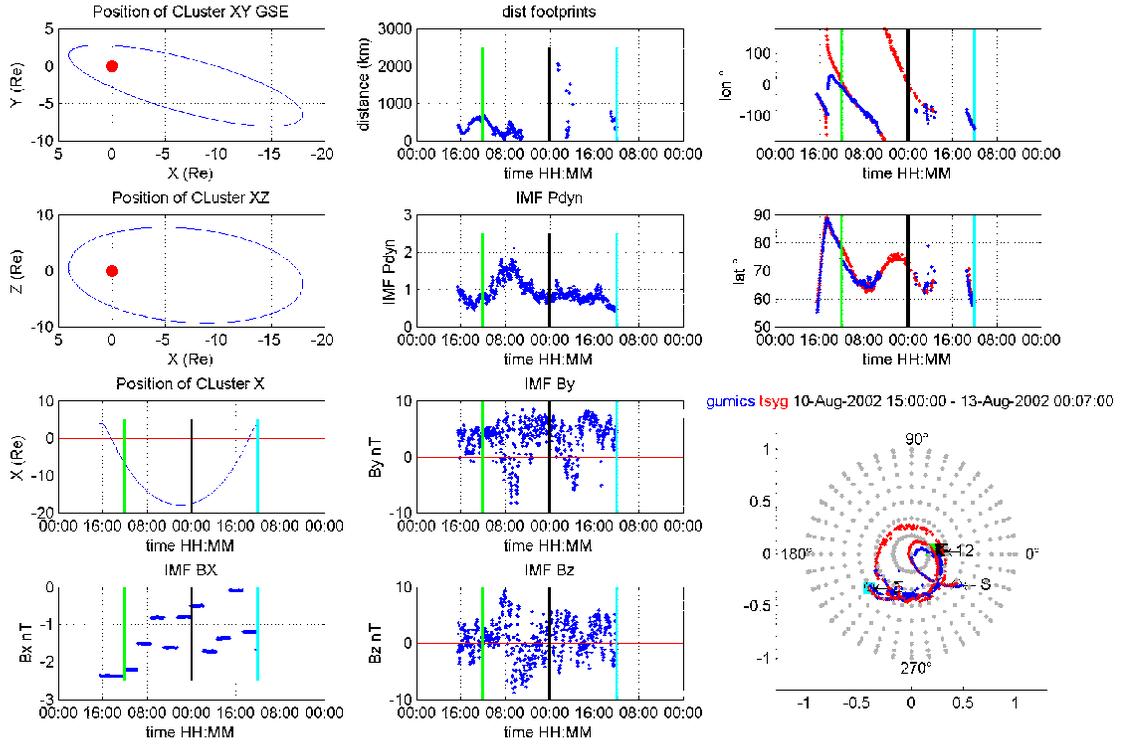}
\caption{Example for August 10\textsuperscript{th}, 2002, northern hemisphere. \textit{1\textsuperscript{st} column 1\textsuperscript{st} row:} the Cluster SC3 orbit in the XY GSE plane. \textit{1\textsuperscript{st} column 2\textsuperscript{nd} row:} the Cluster SC3 orbit in the XZ GSE plane. \textit{1\textsuperscript{st} column 3\textsuperscript{rd} row:} the Cluster SC3 X position. \textit{1\textsuperscript{st} column 4\textsuperscript{th} row:} the $B_x$ magnetic field GSE X component. \textit{2\textsuperscript{nd} column 1\textsuperscript{st} row:} the distance of footprints determined from the GUMICS$-$4 simulations and T96 model. \textit{2\textsuperscript{nd} column 2\textsuperscript{nd} row:} the solar wind dynamic pressure. \textit{2\textsuperscript{nd} column 3\textsuperscript{rd} row:} the Interplanetary Magnetic Field (IMF) GSE Y component. \textit{2\textsuperscript{nd} column 4\textsuperscript{th} row:} the IMF GSE Z component. \textit{3\textsuperscript{rd} column 1\textsuperscript{st} row:} the longitude of the T96 (red) and GUMICS$-$4 (blue) footprints in the SM system. \textit{3\textsuperscript{rd} column 2\textsuperscript{nd} row:} the latitude of the T96 (red) and GUMICS$-$4 (blue) footprints in the SM system. \textit{3\textsuperscript{rd} column 3\textsuperscript{rd}-4\textsuperscript{th} row:} the T96 (red) and GUMICS$-$4 (blue) footprints in SM coordinates.}
\label{fig:aug10}
\end{center}
\end{figure}

\begin{figure}
\begin{center}
\includegraphics[width=0.75\textheight,angle=0]{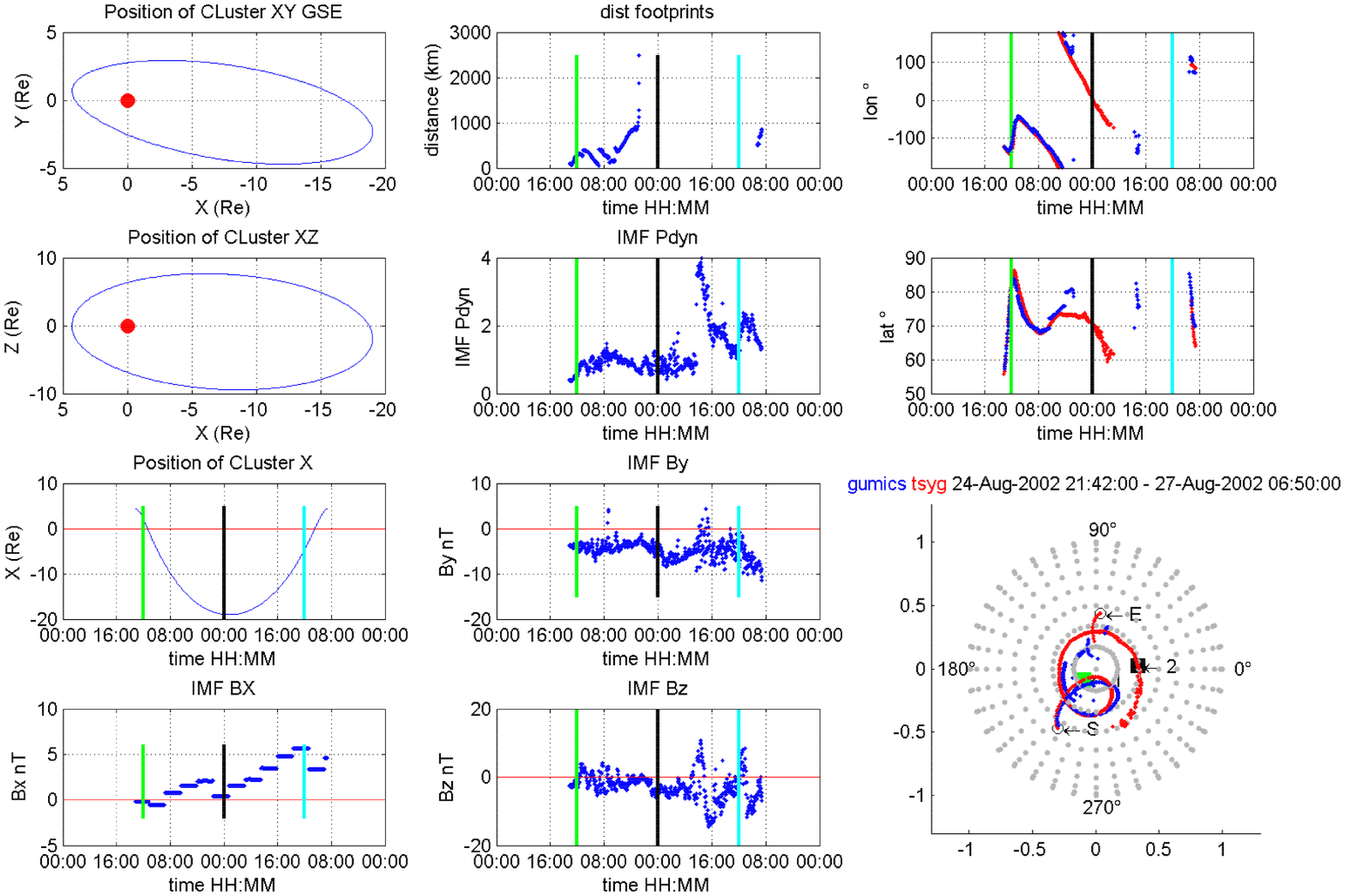}
\caption{Example for August 24\textsuperscript{th}, 2002, northern hemisphere. See Figure~\ref{fig:aug10} for the description of the panels.}
\label{fig:aug24}
\end{center}
\end{figure}

\begin{figure}
\begin{center}
\includegraphics[width=0.75\textheight,angle=0]{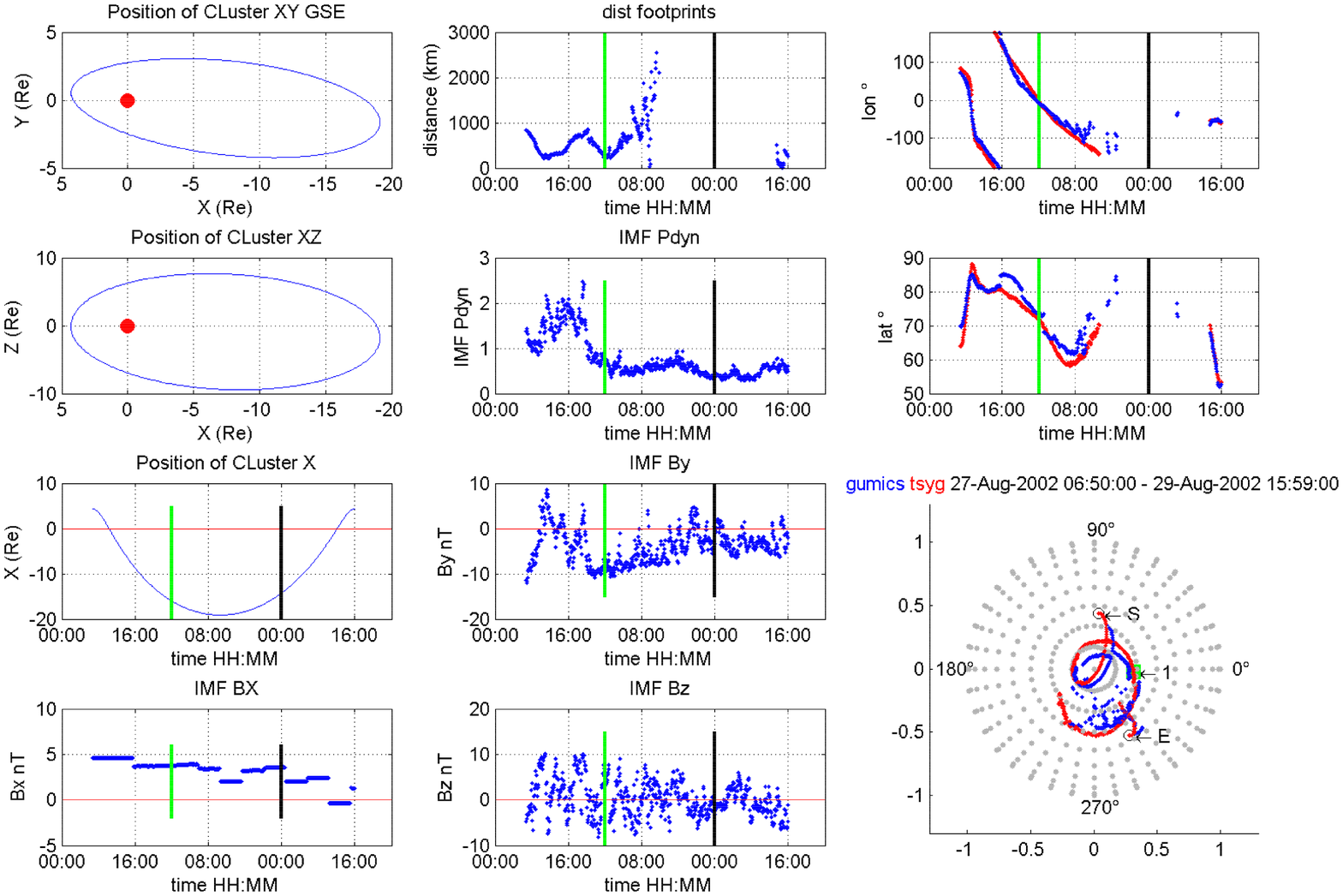}
\caption{Example for August 27\textsuperscript{th}, 2002, northern hemisphere. See Figure~\ref{fig:aug10} for the description of the panels.}
\label{fig:aug27}
\end{center}
\end{figure}

Figure~\ref{fig:feb7} gives an example from February 7th in 2002. This quiet magnetospheric condition case provides an example of an event when the models gives a relatively similar magnetic footprints in the northern and the southern hemispheres. Figure~\ref{fig:aug10}, instead, shows an example from August, 10th, 2002, for the northern hemisphere, where $B_{y}$ is mostly larger than zero. In this case the longitude of the footprints derived from GUMICS$-$4 differ substantially from the footprints derived from T96, as it has been seen in the scatter plot in Figure~\ref{fig:scatter}c,\,d, previously. Figure~\ref{fig:aug24} shows an example for August, 24th, 2002, for the northern hemisphere, where the results of both models are in similar kind of footprints before the Cluster 3 spacecraft had reached apogee. However, at the apogee, presumably because the GUMICS$-$4 model has a shorter magnetotail than T96, differences between the models become larger. Finally, Figure~\ref{fig:aug27} shows an example of August 27th, 2002, for the northern hemisphere, where there is a solar wind pressure pulse. In this case, GUMICS-4 results is substantially different footpoint positions than T96. Furthermore, due to continuous pressure variations, the 4.75 hour sub-run interval in the GUMICS$-$4 year run is becoming obvious as clear steps, also in the ionospheric footprint. The same plots for the southern hemisphere show even less agreement between GUMICS$-$4 and T96, most probably because the GUMICS$-$4 model uses a tilted dipole magnetic field in the ionosphere (not shown).

When the magnetic $B_{y}$ component is different from zero, the longitudes of the footprints for GUMICS$-$4 and T96 deviate considerably from each other, particularly when $B_{y}<0\,nT$. This might be due to the configuration of the magnetotail in the GUMICS$-$4 model. Furthermore, a mismatch between GUMICS$-$4 and T96 footprints arise when the apogee of the Cluster orbit is in the magnetotail, probably due to the shorter magnetotail in the GUMICS$-$4 model. Since GUMICS$-$4 runs are carried out in 12 slices per orbit, 4.75 hour steps can be seen in the footprints as well. This is particularly obvious for varying magnetospheric conditions, e.g., when there is a pressure pulse close to the transition between the 4.75 h slots. However, well within the slots, GUMICS$-$4 responds relatively well to solar wind variability even under disturbed conditions.

\section{Discussion}
\label{sec:disc}

In this study a 368$-$day time period global MHD simulation is launched and analyzed. The GUMICS$-$4 uses only a single processor, therefore the 155 Cluster orbit long time period is divided into 1860 subintervals (slices) and 1860 GUMICS$-$4 simulations are launched. This mandatory technical decision is a potential source for inaccuracy in the GUMICS-4 simulation results. Moreover, another main source of possible inaccuracy are the input parameters. The OMNI solar wind data is derived from other spacecraft measurements. These additional sources of input data inaccuracies -- namely the timeshift to the sub-solar point of the terrestrial bow shock -- increase the risk of the failure of the simulations. It is therefore necessary to use the ACE measurement from the $L_1$ inner Lagrange point. There are more data gaps, however the only calculation is a 20-40\,minute timeshift in the parameters. 

From qualitative comparisons between the GUMICS$-$4 simulation and T96 during a year run, we can conclude that they give relatively similar footprints during quiet conditions for the northern hemisphere. Generally, the matching of the footprint latitude between the GUMICS$-$4 simulation and T96 is reasonably good for all magnetospheric conditions. However, the observed discrepancy is always worse for the southern hemisphere due to the assumed dipole magnetic field in the GUMICS$-$4 simulations. In the future, this hypothesis could be investigated by replacing the GUMICS-4 the simple dipole magnetic field with the IGRF magnetic field model. The step errors at the transition of the sub-run intervals are more difficult to correct, as they arise from the assumption of two hours of steady solar wind for the initialization of every GUMICS$-$4 sub-run and a constant dipole tilt angle in that period. During disturbed solar wind conditions this assumption will introduce a bias to the system, as the real solar wind should be influencing the modelled magnetospheric configuration for the first few minutes of every sub-run, while now there is a constant starting value assumed. Subsequent to the passage of the assumed constant solar wind region towards the deep tail, GUMICS$-$4 again develops a magnetosphere that corresponds to the measured solar wind. Depending on the solar wind speed this might give inaccuracies in the magnetospheric configuration on the day side and the near-Earth region during the first 3 to 10 minutes of every sub-run interval, as seen in the step of the footprint comparison.

The length of the data gaps is the shortest in the selected 368$-$day term period during 2001-2011 (Figure~\ref{fig:gaps}). This choice maximized the length of the simulations. However, the Cluster spacecraft were launched in July and August 2000 and the magnetometers and the plasma instruments were switched off or calibrated frequently in 2002. This produced many data gaps in the Cluster measurements on all spacecraft and limited the accuracy of the comparison of real measurements and simulations. An additional problem is the five$-$minute resolution of the simulation data. There was no data saving capacity to save the simulation status more frequently, however the cross calibration calculation and other methods cannot be applied that efficiently. A forthcoming paper will extend the comparison study for the main regions: the solar wind, the magnetosheath, the day side magnetosphere and the tail. In addition, it would be desirable to compare the magnetic field components and magnitude, the solar wind velocity components and the density in each region. This will be addressed in the follow-up paper. Furthermore, in future, the features of the bow shock, magnetopause and neutral sheet will also be compared in simulations and in Cluster measurements in order to obtained deeper insight into the pros and cons of the MHD approach.  

\section{Summary and conclusions}
\label{sec:concl}

A long global MHD simulation lasting approximately one year (368 days was launched using GUMICS$-$4 code and compared to satellite measurements. The authors knowledge, this is the longest 3D MHD model simulation made so far to make a detailed comparison with observations. The simulation was made based on the previous experience of 162 stationary runs using the same global MHD code. Solar wind data derived from the OMNI was used as simulation input. The 365$-$day long interval that has the shortest data gap during the operation time of the Cluster fleet (2001-2012) was selected for input. Using correlation calculation we proved that the OMNI can be applied to Cluster measurements because the IMF $B_z$ variations are similar to those in the solar wind. The OMNI shifts its solar wind observations to the sub-solar point of the terrestrial bow shock, however this transformation does not overlap the different simulation results. The GUMICS$-$4 typically runs slower than real time, hence we divided the interval of approximately one year into 1860 sub-intervals to complete the simulation faster. This method - simulation in sub-intervals or slices - has no significant influence on the quality of the simulation. 

The Cluster SC3 magnetic footprints were determined in the GUMICS$-$4 simulations. The study showed that the determined footprints were relatively well in agreement with the T96 empirical model, however the footprints agree better in the northern hemisphere than the southern one during quiet conditions. The correlation in latitude does not depend on magnetopsheric conditions. When $B_y$ is non-zero, the correlation between models is worse in longitude in the southern hemisphere. When the Cluster SC3 was situated in the dayside magnetosphere, the deviation between the footprints was small in the northern hemisphere during quiet conditions. In the magnetotail the deviation between the models became larger at the Cluster apogee, possibly because the GUMICS$-$4 magnetotail was shorter than the T96 tail. The study also suggests that GUMICS$-$4 could not model solar wind pressure pulses as realistically as T96. Overall, the study implies that a 3D MHD model can increase our insight into the response of the magnetosphere to solar wind conditions, but the usage of the solar wind input parameters, the adopted technique to perform the runs and analysis of the realism of the simulation results, requires special attention.

\begin{acknowledgments}
The OMNI data were obtained from the GSFC/SPDF OMNI interface at http://omniweb.gsfc.nasa.gov. The authors thank the FGM Team (PI: Chris Carr), the CIS Team (PI: Iannis Dandouras), the WHISPER Team (PI: Jean-Louis Rauch), the PEACE Team (PI: Andrew Fazakerley) and the Cluster Active Archive for providing FGM magnetic field, CIS HIA ion plasma, WHISPER and PEACE electron density measurements. Data analysis was partly done with the QSAS science analysis system provided by the United Kingdom Cluster Science Centre (Imperial College London and Queen Mary, University of London) supported by the Science and Technology Facilities Council (STFC). This research was funded by the European Union Seventh Framework Programme (FP7/2007-2013) under grant agreement No.~263325 (ECLAT) and No.~262863 (IMPEx). The work of Ilja Honkonen is supported by the Academy of Finland and the European Research Council Starting grant number 200141-QuESpace. Laurianne Palin and Hermann Opgenoorth thank the Swedish National Space Board for funding. Eija Tanskanen acknowledges financial support from the Academy of Finland for the ReSoLVE Centre of Excellence (project No.~272157). The work of G{\'a}bor Facsk{\'o} is supported by the OTKA Grant K75640 of the Hungarian Scientific Research Fund. G{\'a}bor Facsk{\'o} thanks Liisa Juusola for the useful discussions and comments; furthermore Anna-M\'aria V\'\i gh and William Martin for improving the English of the paper. The authors thank Pekka Janhunen for developing the GUMICS$-$4 code; as well as the Finnish Meteorological Institute (especially for Lasse Jalava and Markku Hakola) for providing the computer facilities for carrying out the simulations. For further use of the year run data, please contact Minna Palmroth (minna.palmroth@fmi.fi).
\end{acknowledgments}




\pagebreak 

\end{article}


\begin{table}[t]
\caption{The 27 selected intervals in the solar wind. From left to right: the beginning and end of the intervals, the calculated timeshifts of the OMNI vs. the Cluster SC3 $B_z$ magnetic field component cross correlation calculation. OMNI comments: \textsuperscript{1}Data gap in OMNI data.}
\begin{center}
\begin{tabular}{cccc}
\hline
\multicolumn{2}{c}{Interval} & \multicolumn{2}{c}{OMNI vs. Cluster SC3}\\
Start & End & \mbox{Timeshift [min]} & Correlation   \\
\hline
20020201 20:00 & 20020203 04:00 & -1 & 0.96 \\
20020209 01:00 & 20020209 06:00 & 5 & 0.87 \\
20020211 13:00 & 20020212 12:00 & 1 & 0.81 \\
20020213 16:00 & 20020214 08:00 & 10 & 0.83 \\
20020218 09:00 & 20020219 02:00 & -1 & 0.93 \\
20020219 06:30 & 20020219 15:00 & -1 & 0.94 \\
20020220 18:30 & 20020222 00:00 & 1 & 0.84 \\
20020318 17:30 & 20020319 02:30 & -1 & 0.88 \\
20020323 16:00 & 20020323 18:30 & -5 & 0.99 \\
20020412 20:30 & 20020413 02:00 & -2 & 0.93 \\
20020423 16:30 & 20020423 22:00 & -4 & 0.90 \\
20021206 15:30 & 20021206 18:00 & 0 & 0.90 \\
20021229 20:00 & 20021230 16:00 & 0 & 0.63 \\
20030101 16:00 & 20030101 21:00 & -27\textsuperscript{1} & 0.83 \\
20030103 12:00 & 20030104 02:00 & 2 & 0.69 \\
20030106 06:00 & 20030106 19:00 & 2 & 0.76 \\
20030108 07:00 & 20030109 03:30 & 4 & 0.59 \\
20030110 17:00 & 20030110 20:30 & 1 & 0.94 \\
20030113 08:30 & 20030113 18:00 & 0 & 0.91 \\
20030116 02:30 & 20030116 05:30 & 25 & 0.57 \\
20030118 00:00 & 20030118 18:00 & 3 & 0.74 \\
20030120 07:30 & 20030120 13:00 & 1 & 0.80 \\
20030122 12:00 & 20030123 14:00 & 1 & 0.79 \\
20030124 18:00 & 20030126 00:00 & 2 & 0.70 \\
20030127 16:00 & 20030128 06:00 & -3 & 0.87 \\
20030129 12:00 & 20030130 18:00 & 1 & 0.87 \\
20030203 06:00 & 20030204 00:00 & 4 & 0.61 \\
\hline
\end{tabular}
\end{center}
\label{tab:omniclcorr}
\end{table}

\end{document}